# All-optical pulse compression of broadband microwave signal based on stimulated Brillouin scattering

**Xin Long,[1] Weiwen Zou,[1,2,*] and Jianping Chen[1,2]**

[1]*State Key Laboratory of Advanced Optical Communication Systems and Networks, Department of Electronic Engineering, Shanghai Jiao Tong University, Shanghai 200240, China*
[2]*Shanghai Key Lab of Navigation and Location Services, Shanghai Jiao Tong University, Shanghai 200240, China*
[*]*wzou@sjtu.edu.cn*

**Abstract:** Pulse compression processing based on stimulated Brillouin scattering (SBS) in an optical fiber is theoretically and experimentally demonstrated. Broadband microwave signal is electro-optically modulated onto the pump lightwave that is launched into one end of the fiber. Acoustic wave in the fiber inherits the amplitude and phase information of the pump lightwave and thus the coupling between the acoustic wave and pump lightwave leads to the auto-correlated process of the pump lightwave as well as the modulated microwave signal. Derivation of the SBS coupling equations shows that the short-pulse probe lightwave amplified by the pump lightwave possesses the nature of auto-correlation formula. All-optical pulse compression of the broadband microwave signal is implemented after a subtraction between the detected probe pulse with and without SBS. A proof-of-concept experiment is carried out. The pulse compression of a linear frequency-modulated microwave signal with 1 GHz sweep range at the carrier frequency of 4.3 GHz is successfully realized, which well matches the theoretical analysis.

## 1. Introduction

Pulse compression raised up in 1960s is a key signal processing technique to realize better spatial resolution and higher signal-to-noise ratio in the radar field [1]. It is in principle a process that the received broadband signal correlates with the conjugated signal. Many efforts have been made to improve the devices and waveforms in pulse compression system [2]. Among them are the digital processing and surface-wave processing that allow the implementation of more complicated signal waveforms [3-5]. As modern coherent radar systems utilize ultrahigh frequency signals with much larger bandwidth, the signal processing and phase stability have become significant issues to be concerned [6]. For instance, the matched filtering process of the microwave signal for modern coherent radar application is always required after the digitization [1, 2]. It leads to a big challenge for analog-to-digital conversion (ADC) that should provide sufficient analogue bandwidth, ultra-high sampling rate, and great sampling resolution (also called the effective number of bits) [7]. The pulse duration of the radar's microwave signal is sufficiently long so as to increase its transition distance, which sharply increases the depth and speed difficulty of digital storage. Therefore, the huge bit-rates of the digitized signal due to the broad bandwidth and long pulse duration of the microwave signal make ADC, digital storage, and signal processing more and more challengeable.

Microwave photonics [8-10] combines the merits in microwave and photonic techniques and enables the generation [11-13], control [14], and processing [15-17] of broadband microwave signals with high phase stability. To overcome the bandwidth limit of the electronic ADC, several types of photonic ADCs [18-20] have been proposed. However, photonics ADCs can't reduce the requirement of digital storage and signal processing. Thus, an all-optical scheme that achieves auto-correlation process before digitization is significantly desired. To the best of our knowledge, the analogue matched filtering for the pulse compression of the broadband microwave signal has not been reported so far. In optical fibers, stimulated Brillouin scattering (SBS) is in principle three-wave interaction among the pump lightwave, the counter-propagated probe lightwave, and the acoustic wave [21]. The acoustic wave is excited due to the electro-striction effect and inherits the amplitude and phase information from the optical waves [22]. Such a property has been successfully utilized in other applications to fiber-optic sensors [23, 24] and slow light [25-27].

In this paper, we demonstrate the SBS-based pulse compression for all-optical characterization of broadband microwave signals. As mentioned above, acoustic wave inherits the amplitude and phase information of the signal to be processed [22], hence the time delay and amplitude conjugation conditions that are necessary for the matched filtering are automatically satisfied during the SBS process. The pump lightwave carrying the microwave signal amplifies the short-pulse probe lightwave via the acoustic wave. Consequently, the pulse compression is implemented by direct detection of the probe pulse with and without SBS, which is short in time and thus significantly reduces the difficulty of digital storage and signal processing. Experiment is conducted to verify the feasibility of the pulse compression of linear frequency modulated (LFM) pulses with 1 μs pulse duration and 1 GHz sweep range at 4.3 GHz carrier frequency. The range resolution of compressed LFM signal is 0.88 ns, which is in good agreement with the ideal auto-correlation of the LFM pulse.

## 2. Principle

Without loss of generality, we assume that the pump and probe lightwaves are launched into the two ends of an optical fiber at $z = 0$ (near end) and $z = L$ (far end), respectively. $L$ denotes the length of the fiber. The coupling equations describing the SBS interaction in the fiber is given by [21]:

$$\left(\frac{\partial}{\partial t}+\Gamma\right)Q(z,t) = -ig_1 E_P(z,t) E_S^*(z,t), \tag{1}$$

$$\left(\frac{\partial}{\partial z}+\frac{1}{v}\frac{\partial}{\partial t}\right)E_P(z,t) = ig_2 E_S(z,t) Q(z,t), \tag{2}$$

$$\left(\frac{\partial}{\partial z}-\frac{1}{v}\frac{\partial}{\partial t}\right)E_S(z,t) = -ig_2 E_P(z,t) Q^*(z,t), \tag{3}$$

where $g_1$ and $g_2$ are electro-strictive parameters. $v$ denotes the light velocity in the fiber. $Q(z,t)$, $E_P(z,t)$ and $E_S(z,t)$ represent slowly-varying envelopes of the acoustic wave, pump and probe fields at position $z$ and time $t$, respectively. $\Gamma = 1/(2\tau_p) + i(\Delta\omega-\omega_B)$ is the detuning and damping coefficient with $\tau_p$ representing the phonon lifetime and $\Delta\omega$ corresponding to beat frequency between pump and probe. $\omega_B$ is the Brillouin frequency shift (BFS).

Eq. (2) and Eq. (3) can be modified as homogeneous when weak probe lightwave is injected, so that both the pump and probe fields can be expressed by:

$$E_P(z,t) = E_{P0}\left(t-\frac{z}{v}\right), \tag{4}$$

$$E_S(z,t) = E_{S0}\left(t-\frac{L-z}{v}\right), \tag{5}$$

where $E_{P0}(t)$ and $E_{S0}(t)$ denote the complex envelope function for the injected pump and probe lightwaves, respectively.

According to Eq. (4) and Eq. (5), the solution to Eq. (1) is given by:

$$Q(z,t) = -ig_2 \int_{-\infty}^{t} E_{P0}\left(\tau - \frac{z}{v}\right) E_{S0}^{*}\left(\tau - \frac{L-z}{v}\right) e^{-\Gamma(t-\tau)} d\tau. \tag{6}$$

The SBS process and the excitation of the acoustic wave are schematically illustrated in Fig. 1. All the symbols defined above are presented in Fig. 1(a). As shown in Fig. 1(b), an LFM pulse is modulated onto the pump lightwave ($E_P$) at $z = 0$ and the probe lightwave at $z = L$ is a short pulse ($E_S$). The probe pulse is downshifted in frequency from the pump lightwave and the downshifted frequency is equal to the BFS in the fiber. Since the modulated pump lightwave contains a broadband LFM pulse, the bandwidth of the probe lightwave should be wide enough. When they meet each other, the acoustic wave ($Q$) is generated at the same frequency as the BFS [see Fig. 1(c)]. As time goes by, 'new' phonons are excited along the fiber while 'old' ones attenuate and disappear rapidly [see Fig. 1(d)]. Finally, as depicted in Fig. 1(d), the amplified probe lightwave ($E_S$') is detected at $z = 0$ [see Fig. 1(e)].

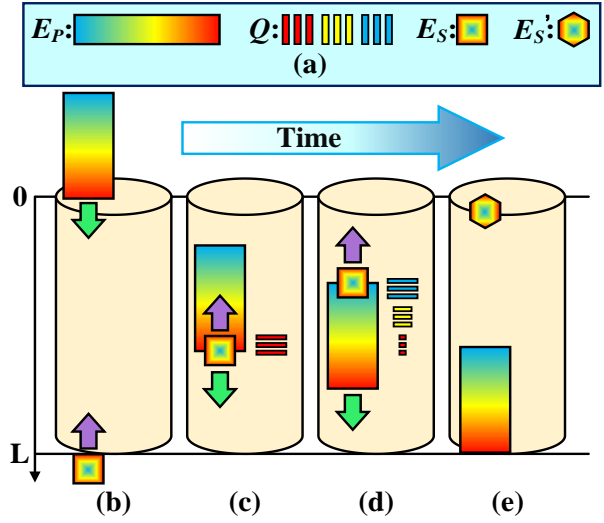

Fig. 1. The SBS process and the excitation of the acoustic wave. (a) Illustration for the symbols. $E_P$: pump lightwave, which is modulated by an LFM pulse. $Q$: acoustic wave. $E_S$: short-pulse probe lightwave. $E_S$': received probe pulse. (b) Pump lightwave is injected at $z = 0$ while the counter-propagating probe lightwave is launched at $z = L$. (c) The first time when the two lightwaves meet each other and acoustic wave is generated. (d) 'New' acoustic waves are stimulated while 'old' ones last for a lifetime and disappear finally. (e) Amplified probe lightwave is detected at the near end of the fiber.

Substituting Eqs. (4-6) into Eq. (3), the electric field of the detected probe lightwave, $E_S$'$(t)$, can be expressed by:

$$E_{S}'(t+T) = \frac{g_B v}{8\tau_p}\left\{E_{S0}(t) \otimes \left[u(t) y_P(t) e^{-\Gamma^* t}\right]\right\} + E_{S0}(t), \tag{7}$$

where $g_B = 4g_1 g_2 \tau_p$ is the Brillouin gain coefficient, $T = L/v$ represents the period needed for optical light to travel through the fiber, the symbol of $\otimes$ denotes the convolution algebra, and $u(t)$ is the unit step function. $y_P(t)$ in Eq. (7) is given by:

$$y_P(t) = E_{P0}(t) \otimes E_{P0}^{*}(-t), \tag{8}$$

which corresponds to the auto-correlation formula of the pump field $E_{P0}(t)$ that is modulated by the broadband microwave signal.

Note that $E_S$'$(t+T)$ and $E_{S0}(t)$ represent the probe field at near end with and without the pump lightwave running into the fiber, their subtraction is needed to extract the term of $y_P(t)$ from $E_S'(t+T)$. For a direct detection system, the detected voltage signal is proportional to the optical power of the probe lightwave. Thus the signal after subtraction can be expressed as follows:

$$s(t+T) \propto \left|E_S'(t+T)\right|^2 - \left|E_{S0}(t)\right|^2 = \left|\Delta E_S(t)\right|^2 + 2\,\mathrm{Re}\left\{\Delta E_S(t) E_{S0}^*(t)\right\}, \tag{9}$$

where $\Delta E_S(t) = E_S'(t+T) - E_{S0}(t)$ represents the Brillouin gain factor and $Re\{\cdot\}$ denotes real part of the complex signal. $s(t+T)$ is divided into two parts in terms of different time. When $t$ is smaller than the probe pulse duration $\tau_S$, Brillouin gain can be considered as a small signal and thus the term of $|\Delta E_S(t)|^2$ in Eq. (9) can be ignored [see Eq. (10)]. Whereas $t$ is larger than $\tau_S$, the term of $2Re\{\Delta E_S(t) E_{S0}^*(t)\}$ can be neglected as a result of $E_{S0}(t) \approx 0$ for a high extinction ratio (ER) of the short-pulse probe lightwave [see Eq. (11)].

$$s(t+T) \propto 2\,\mathrm{Re}\left\{\Delta E_S(t) E_{S0}^*(t)\right\}, \quad \text{when } t < d_S, \tag{10}$$

$$s(t+T) \propto \left|\Delta E_S(t)\right|^2, \quad \text{when } t \geq d_S, \tag{11}$$

Simulation result is depicted in Fig. 2. A 1 μs LFM pulse with 1 GHz sweep range is modulated onto a continuous-wave pump lightwave, and a 0.5 ns pulse with 40 dB ER is used as the probe lightwave. The BFS is set uniform along the fiber and $\tau_p = 10$ ns. Figure 2(a) shows the comparison of the detected probe power with and without the pump lightwave. The above-mentioned two parts in the amplified probe power can be identified. One corresponding to $t < \tau_S$ is accumulated from the probe lightwave and is not useful for this study. Whereas the other at the rear of the pulse corresponds to the desired pulse compression of the LFM pulse. After the horizontal coordinate adjustment in Fig. 2(a), these two parts are separated by $t = 0$. Figure 2(b) shows the subtracted signal defined in Eq. (9). It is compared with the small signal approximation given by the term of $|\Delta E_S(t)|^2$ in Eq. (11) and the squared ideal pulse compression of the LFM pulse $|y_P(t)|^2$ defined in Eq. (8). The ideal pulse compression of the broadband microwave signal matches well the small signal approximation as well as the directly subtracted power. For an LFM pulse with large time-wavelength product, the range resolution ($R$) for its pulse compression is defined as the width between the peak and the first null, which approximately equals to the reciprocal of the bandwidth (*B*) of the LFM pulse, as follows [28]:

$$R \approx \frac{1}{B}. \tag{12}$$

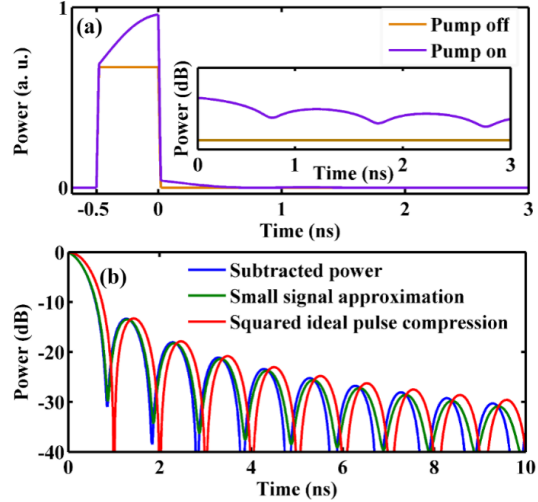

Fig. 2. Simulation results for the pulse compression based on the SBS process. The pump lightwave is modulated by an LFM pulse with 1 GHz sweep range and the probe lightwave is a short pulse with 0.5 ns duration. (a) The detected probe power with and without the pump lightwave. The inset shows the log scale of the vertical axis. (b) The subtracted power is compared to the small signal approximation and the squared ideal pulse compression.

As we can see from Fig. 2(b), the range resolutions *R* for the small signal approximation and the subtracted power equal to each other, that is, 0.86 ns, which approximately equals to that of the ideal pulse compression (1 ns). The possible attribution to the tiny difference will be demonstrated in the Discussion session.

## 3. Experimental details

Figure 3 depicts the experimental setup. Light from a 1550 nm distributed-feedback laser (DFB-LD, NEL NLK1C6DAAA) is split into two paths by a 1:1 fiber coupler. The upper branch is used as the pump lightwave. Through an electro-optic modulator (EOM1), the pump lightwave is modulated by the input microwave signal to be processed. The lower branch is modulated by another EOM2 (Eospace, AX-6K5-10-PFU-PFUP-R4). To generate the frequency difference between these two branches, a single sideband modulator (SSBM) is additionally used in the lower branch. The first lower sideband of the modulated signal after SSBM is selected and thus the beat frequency between the pump and probe lightwaves equals to the BFS of the optical fiber. Polarization controllers (PC1, PC2, and PC3) are used to optimize the light polarizations before the modulators. Erbium-doped fiber amplifiers (EDFA1 and EDFA2) are utilized to compensate and control the optical power of the pump and probe lightwaves. PC4 and PC5 ensure the maximum SBS interaction in a standard single-mode optical fiber (SMF). A circulator is used for the isolation of the pump lightwave and transmission of the amplified probe lightwave. A photo-detector (PD) converts the detected probe power into the electric signal, namely the compressed signal.

As a proof-of-concept experimental verification, the pump lightwave is modulated by an LFM pulse which is generated via a voltage-controlled oscillator (VCO, Mini-circuits ZX95-5400-S+) and an RF amplifier (Mini-circuits ZX60-542LN-S+). The LFM pulse's duration is set to be 1 μs and its frequency is linearly swept. Since the frequency of the pump lightwave is linearly modulated, a 0.5 ns pulse with sufficient bandwidth is modulated on the probe lightwave so that the beat frequency between the pump and probe lightwaves meeting at different positions of the 200-meter-long SMF maintains to cover 10.813 GHz (i.e. the BFS of the SMF). The driving frequency for the SSBM is set to be 6.513 GHz. The average power of the pump and probe lightwaves before entering the SMF are 20 dBm and 10 dBm, respectively. Each experimental result is acquired after 1000-times average.

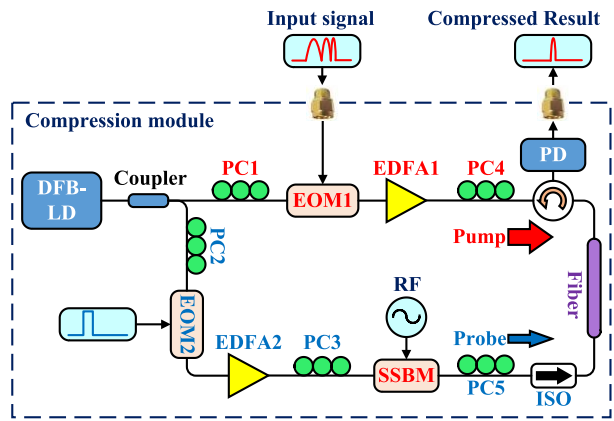

Fig. 3. Experimental setup for pulse compression of the broadband microwave frequency based on SBS process. DFB-LD: distributed-feedback laser. PC: polarization controller. EOM: electro-optic modulator. EDFA: erbium-doped fiber amplifier. ISO: isolator. SSBM: single sideband modulator. PD: photo-detector.

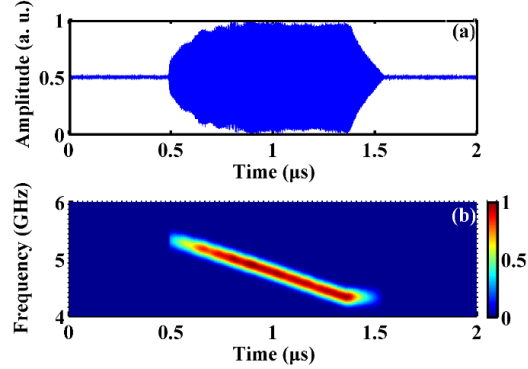

Fig. 4. Magnitude (a) and short-time Fourier transform (b) of the utilized LFM pulse with 1 μs duration and 1 GHz sweep range.

Figure 4 presents the magnitude of the LFM pulse [Fig. 4(a)] and its corresponding short-time Fourier transform result [Fig. 4(b)]. It is shown that the pulse duration of the LFM is 1 μs and its frequency sweeps linearly from around 5.3 GHz to 4.3 GHz. The bandwidth-to-carrier ratio of the LFM pulse is 23.3 %.

By turning on and off the EDFA1, the detected probe powers with and without the Brillouin gain are obtained and illustrated in Fig. 5(a). The subtracted voltage of these two powers at the rear of the pulse is proportional to the term of $|\Delta E_S(t)|^2$ that is defined by Eq. (9). In comparison, both the squared ideal pulse compression of the measured LFM shown in Fig. 4 and the subtraction of the two probe powers in Fig. 5(a) are shown in Fig. 5(b). The range resolution of the experimental curve is about 0.88 ns, which agrees well with the theoretical value shown in Fig. 2(b) whereas a little bit smaller than the reciprocal of 1 GHz sweep range (1 ns). The physical reason will be also presented in Discussion section. Moreover, the range resolution $R$ (or compressed width) for the LFM pulse with different $B$ is depicted in Fig. 6. It shows a clear inverse relation between $R$ and $B$ of the LFM pulse, which matches well Eq. (12).

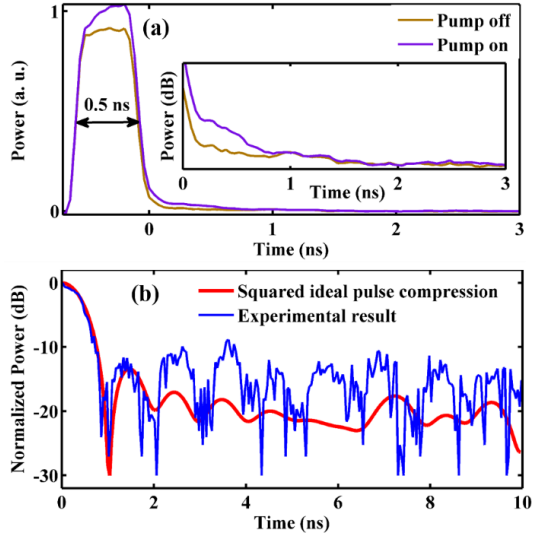

Fig. 5. Experiment results for the pulse compression based on the SBS process. (a) The detected probe power with and without the pump lightwave injected. (b) The pulse compression achieved by the subtraction in (a) is compared with the squared ideal pulse compression of the LFM pulse shown in Fig. 4.

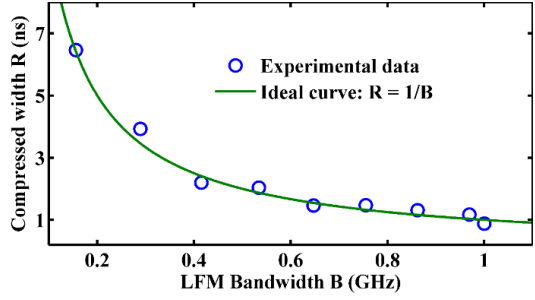

Fig. 6. The full width of the compressed mainlobe as a function of the sweeping bandwidth of the LFM pulse.

## 4. Discussion

As can be seen from Fig. 2(b) and Fig. 5(b), there are more or less mismatches in the simulation and experiment. Comparison among Eq. (7), Eq. (8), and Eq. (9) shows that the term of $\Delta E_S(t)$ is just an approximation of $y_P(t)$. When $t > 0$, $\Delta E_S(t)$ is exponentially reduced and the condition of $|\Delta E_S(t)|^2 \gg 2Re\{\Delta E_S(t)E_{S0}^*(t)\}$ is somehow broken so that the term $2Re\{\Delta E_S(t)E_{S0}^*(t)\}$ in Eq. (9) cannot be ignored. To further overcome this influence, a short-pulse probe lightwave with much higher ER might work better. In the experiment, although the profiles of the sidelobes can be roughly observed [see Fig. 5(b)], the noise of the photo-detector and laser diode distorts the detected probe power and thus influences the compressed pulse (especially, the weak sidelobes).

In Eq. (7), the convolution between $E_{S0}(t)$ and $u(t)y_P(t)\exp(-\Gamma^* t)$ brings in distortion to the compressed term of $|\Delta E_S(t)|^2$. The probe pulse duration $\tau_s$ is one of the key parameters that determines this distortion. Figure 7(a) shows the simulation result of $|\Delta E_S(t)|^2$ with different values of $\tau_s$. The sweep range of the LFM pulse is 1 GHz and $\tau_p = 10$ ns. When $\tau_s$ increases, the mainlobes and sidelobes of $|\Delta E_S(t)|^2$ become narrow and the dips among the sidelobes gradually disappear. It attributed to the discrepancy in the mainlobe of Fig. 5(b). Better experimental result of the all-optical pulse compression is expectable if a 0.1 ns short-pulse probe lightwave is used.

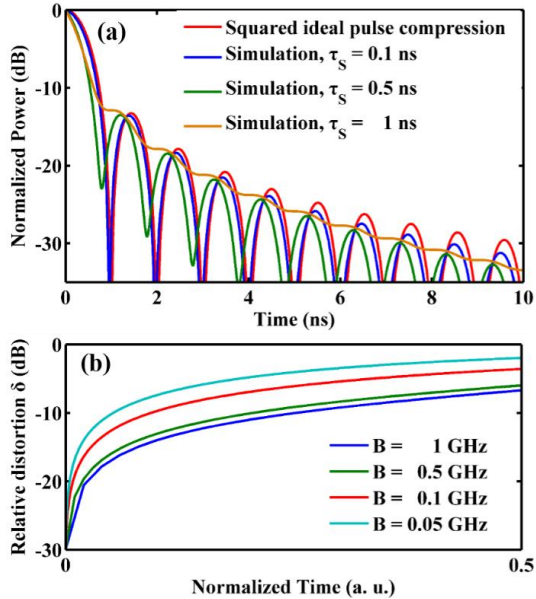

Fig. 7. Analysis on the distortions that induced by the pulse duration ($\tau_S$) of the probe lightwave and the phonon lifetime. (a) Comparison between the squared pulse compression of $|y_P(t)|^2$ and subtracted signal of $|\Delta E_S(t)|^2$ with different $\tau_S$. (b) Relative distortions caused by the phonon lifetime $\tau_p$ for different LFM pulse bandwidths $B$. Horizontal coordinate is normalized by the factor of $1/B$.

Through further analyzing the term of $u(t)y_P(t)\exp(-\Gamma^*t)$ in Eq. (7), the phonon lifetime $\tau_p$ (which is usually 10 ns) also influences the performance of the all-optical pulse compression. To characterize this influence, the relative distortion is defined as follows:

$$\delta = \left[ \left|y_P(t)\right|^2 - \left|\Delta E_S(t)\right|^2 \right] / \left|y_P(t)\right|^2 . \tag{13}$$

Figure 7(b) depicts the simulation results of the relative distortion as a function of the bandwidth (*B*) of the LFM pulse. A 0.1 ns probe pulse with infinite ER $\tau_p = 10$ ns is assumed. The horizontal coordinate (time) is normalized by the factor of $1/B$ so as to compare the relative distortion among all the compressed mainlobes of different LFM pulses. It is shown that the LFM pulses with higher *B* suffer less distortion to the pulse compression. In other words, the all-optical pulse compression proposed in this work is essentially advantageous for broadband microwave signal.

## 5. Conclusion

We theoretically and experimentally demonstrate the SBS based pulse compression for a broadband microwave signal. During the interaction between the pump lightwave and acoustic wave in the SBS process, the time delay and amplitude conjugation conditions for matched filtering are automatically satisfied. This is because the acoustic wave inherits the amplitude and phase information of the microwave signal that is modulated onto the pump lightwave. Theoretical derivation indicates that the auto-correlation of the microwave signal appears in the formula of the amplified probe field. Thus, all-optical pulse compression is implemented by subtraction of the detected probe lightwave with the pump lightwave from that without pump lightwave. Under the small signal approximation, the subtraction equals to the squared pulse compression of the microwave signal. A proof-of-concept experiment is carried out to validate the pulse compression of an LFM pulse with 1 μs pulse duration and 1 GHz sweep range. The resolution of compressed signal of the broadband microwave signal via the SBS based pulse compression is about 0.88 ns, which is in a reasonable agreement with the auto-correlation of the LFM pulse. Moreover, factors that introduce the discrepancies between the SBS based pulse compression and the ideal one are analyzed, which points out the future efforts to avoid the distortions that caused by the noise, pulse duration, and phonon lifetime.

**Acknowledgments**

This work was partially supported by National Natural Science Foundation of China (Grant Nos. 61571292 and 61535006) and by the State Key Lab Project of Shanghai Jiao Tong University (Grant Nos. 2014ZZ03016).